\documentclass[manuscript,screen]{acmart}

\usepackage[normalem]{ulem} 
\AtBeginDocument{%
  \providecommand\BibTeX{{%
    \normalfont B\kern-0.5em{\scshape i\kern-0.25em b}\kern-0.8em\TeX}}}

\setcopyright{acmcopyright}
\copyrightyear{2022}
\acmYear{2022}
\acmDOI{XXXXXXX.XXXXXXX}

\begin{document}


\title{ANONYMOUS VOTING SCHEME USING QUANTUM ASSISTED BLOCKCHAIN}

\author{Sandeep Mishra}
\email{sandeep.mtec@gmail.com}
\orcid{0000-0002-5190-9754}
\affiliation{%
  \institution{Jaypee Institute of Information Technology}
  \streetaddress{A-10, Sector-62}
  \city{Noida}
  \state{UP}
  \country{India}
  \postcode{201309}
}

\author{Kishore Thapliyal}
\email{kishore.thapliyal@upol.cz}
\orcid{0000-0002-4477-6041}
\affiliation{%
  \institution{Joint Laboratory of Optics of Palack\'{y} University and Institute of Physics of CAS, Faculty of Science, Palack\'{y} University}
  \streetaddress{17. listopadu 12, 771 46}
  \city{Olomouc}
  \country{Czech Republic}}

\author{S Krish Rewanth}
\affiliation{%
  \institution{Indian Institute of Information Technology, Design and Manufacturing}
  \city{Kancheepuram}
  \country{India}
}

\author{Abhishek Parakh}
\email{aparakh@unomaha.edu}
\affiliation{%
 \institution{University of Nebraska}
  \city{Omaha}
 \country{USA}}

\author{Anirban Pathak}
\email{anirban.pathak@gmail.com}
\orcid{0000-0003-4195-2588}
\affiliation{%
  \institution{Jaypee Institute of Information Technology}
  \streetaddress{A-10, Sector-62}
  \city{Noida}
  \state{UP}
  \country{India}
  \postcode{201309}
}

\renewcommand{\shortauthors}{Mishra et al.}

\begin{abstract}
  Voting forms the most important tool for arriving at a decision in any institution. The changing needs of the civilization currently demands a practical yet secure electronic voting system, but any flaw related to the applied voting technology can lead to tampering of the results with the malicious outcomes. Currently, blockchain technology due to its transparent structure forms an emerging area of investigation for the development of voting systems with a far greater security. However, various apprehensions are yet to be conclusively resolved before using blockchain in high stakes elections. Other than this, the blockchain based voting systems are vulnerable to possible attacks by upcoming noisy intermediate scale quantum (NISQ) computer. To circumvent, most of these limitations, in this work, we propose an anonymous voting scheme based on quantum assisted blockchain by enhancing  the advantages  offered by blockchain with the quantum resources such as quantum random number generators and quantum key distribution. The purposed scheme is shown to satisfy the requirements of a good voting scheme. Further, the voting scheme is auditable and can be implemented using the currently available technology.
\end{abstract}




\maketitle

\section{Introduction}

In modern societies, voting is considered as the best way for arriving at a decision which involves many stakeholders. Often people use open voting to decide on trivial issues e.g., deciding the  place to visit for a family holiday, going to a restaurant, deciding player of the match in sports. But, there are many situations where people want to know the outcome of voting, but the voters do not want their votes to be identified. Such voting protocols are  anonymous voting protocols in which the link between the voter and the vote is hidden. In fact, anonymous voting system helps the voter to exercise the vote without any fear, pressure or any future repercussions. The easiest way to implement anonymous voting is via the use of paper ballot in which every voter casts the vote by putting their choice in the ballot box. After every voter has voted, the ballot box is opened and results are announced after tallying the votes. Even though the paper ballot system is considered  the best way to implement anonymous voting, the changing needs of the society demands an alternate system which is at least as good as paper ballot system. The main disadvantage of paper ballot system is that it requires the voter to visit the voting booths in person to cast the vote.  However, for the society which is connected together via use of technology, this traditional system of voting seems at odds with the current times. Just as people are getting familiar with and addicted to online purchase, banking, meetings, teaching, gaming, etc., the society urgently requires an electronic voting scheme \cite{kohno2004analysis,qadah2007electronic,shahzad2019trustworthy} which has almost similar levels of security as that of the paper ballot based systems.  Further, the world has been affected by lockdowns due to the COVID-19 pandemic which has led to a heightened interest towards the development of secure electronic voting systems \cite{landman2020pandemic,essex2020secure}. Before going further, let us just highlight the minimum requirements \cite{schneier1996,park2021} that any anonymous voting protocol (existing or new) should satisfy.
\begin{itemize}

\item \textbf{Eligibility:} Only eligible voters should be allowed to vote.

\item \textbf{Anonymity:} No one other than voter should know about the voting choice. 

\item \textbf{Non-reusability:} Each voter can vote only once.

\item \textbf{Binding:} No one can change the vote after it has been cast.

\item \textbf{Verifiability:} Each voter should be satisfied that his vote has been recorded  properly.

\item \textbf{Auditable:} Voting process should be such that one is able to  verify the correctness of the outcome of the voting process.

\end{itemize}

The currently used electronic voting schemes are not full-proof and lack many of the above mentioned properties. Further, the possibility of remote voting comes with their own set of challenges for the designing of such electronic voting schemes. Some of the major challenges \cite{park2021} are as follows:
\begin{itemize}

\item \textbf{Authentication:} If the voter is voting from a remote location, one has to verify the credentials to distinguish between a genuine voter  and an impersonator. In fact, the problem of providing a full-proof mechanism for authentication of a genuine voter is of utmost importance in elections where the stakes are too high. Further, even the authenticated voter should not be allowed to vote more than once.

\item \textbf{Coercion free:} One has to ensure that voter is voting as per his free will without any external fear or pressure. 

\item \textbf{Receipt free:} Voting process should not lead to trail via which the voter could prove his choice of vote to anyone. Further, the process should not  reveal the voter's choice of vote.

\item \textbf{Auditable:}   The voting authority should conform to mechanism that allows a voter to inspect whether his vote has been recorded properly and counted. Further, such complying system would provide a course of action to a person or a group who wants to challenge the outcome of the voting.

\end{itemize}

Other than the above aspects, the credibility of most of the currently used electronics based systems as well as paper ballot based voting schemes depends on the trusted central authority  responsible for the  orderly conduct of elections. This trusted authority is responsible for all the tasks, such as voter registration, casting of votes and the tallying phase. Since the records are kept with the central authority, data maybe subjected to deliberate or accidental damage. Therefore, the storage and tallying of the votes in a voting system with central authority is far from being perfect due to feasibility of deliberate manipulations from an adversary. Further, there is opaqueness with regards to the verifiability of the votes as the voter is not completely sure about the methods being used to record the votes which may not be tamper-proof.  Thus, one requires an electronic voting system which is transparent  and do not require a trusted central authority. 

In the recent years, blockchain technology has emerged as a principal tool to create a distributed tamper-proof database with the prime example being the emergence of cryptocurrency called Bitcoin \cite{nakamoto2008bitcoin}. Since then, this technology has found various applications in the fields of finance, healthcare, logistics, and so on \cite{bodkhe2020blockchain,catalini2020some,holbl2018systematic}. The possible use of blockchain technology in voting schemes provides a way for the effective implementation of the electronic voting with far greater transparency. Basically, a blockchain is  a ledger or database which is distributed  among a large number of mutually untrusting parties, and these parties try to  reach at a consensus on the contents of the database \cite{yaga2019blockchain,zheng2017blockchain}. The most interesting thing about blockchain is that there is no central authority with all the parties having a stake over the database. This leads to complete transparency over the addition or deletion of data in the database. Further, all the information is stored in the form of blocks which are linked to each other via the cryptographic hash functions with each new block containing the hash of the previous block. Therefore, if one wants to do any manipulation of data in just one of the blocks then he/she has to modify all the subsequent blocks which are linked to this. Further, every stakeholder can in principle hold a copy of the blockchain with themselves and verify the records of the database. This helps in protection of the data from any unauthorized modification or manipulation. Recently, various protocols for electronic voting using blockchains have been proposed by researchers \cite{wang2018large,abuidris2019survey,sun2019simple,osgood2016future,yi2019securing,
hanifatunnisa2017blockchain,tacs2020systematic,poniszewska2020auditable,huang2021application}. 

Even though blockchains have  several advantages, the security \cite{zhang2019security} is heavily dependent on the classical cryptographic schemes (based on the mathematical complexity). The two main ingredients of the classical  blockchain technology are cryptographic hash functions and digital signatures (based on public key cryptography) which are prone to quantum attacks \cite{aggarwal2017quantum,fedorov2018quantum,kearney2021vulnerability}. It has been shown by Shor \cite{shor1994,shor1999} that  some famous classical public key cryptography  schemes can be broken by the application of quantum computers. Further, the Grovers's search algorithm provides a speed up for the calculation of inverse hash function (breaking of hash) \cite{grover1996fast}. In addition, if an adversary controls more than 50 percent of the computing power used for mining, then the contents of blockchain can be compromised \cite{sayeed2019assessing}. Note that, if one has quantum resources in hand then it may be possible to tamper with the blockchain. Thus, the potential advent of a possible quantum computer in the future can create a havoc for the classical blockchain \cite{fedorov2018quantum} and will surely limit its usefulness for application in electronic voting schemes. Recently, some modifications have been proposed in the current blockchain to develop quantum-proof blockchain via use of post-quantum cryptography \cite{bernstein2017post,regev2009lattices,jao2011towards,matsumoto1988public,mceliece1978public,fernandez2020towards} (which are expected to be computationally secure) as well as use of unconditionally secure quantum cryptographic protocols \cite{kiktenko2018quantum,li2019quantum,rajan2019quantum,gao2020novel,banerjee2020quantum}. 

 In the next section, we review the blockchain technology in detail and the possibilities of extending quantum features for the implementation of quantum assisted blockchain. Our aim is to discuss the features of blockchain at a higher level and omit the detailed structure and mechanism required for its realistic implementation. Thereafter taking into account the possible advantages offered by quantum assisted blockchains, in the subsequent sections, we theoretically develop a secure and realistic blockchain based electronic voting protocol. This scheme satisfies all the requirements of voting and is free from the possible future attacks due to the quantum computers.

 \section{Blockchain and Quantum assisted blockchain}

Blockchain technology came into picture in 2008 with the introduction of cryptocurrency Bitcoin \cite{nakamoto2008bitcoin} by a pseudo-name Satoshi Nakamoto. The main feature of Bitcoin is that it is a decentralized currency and has all the features for controlled minting of coins along with a validity check of all the transactions in spite of it being a public record. Since then, there has been a continuous increase in the  launch of new crypto currencies, such as Ethereum \cite{ethereum2014}, Litecoin, Bitcoin Cash, Monero and many more \cite{giudici2020cryptocurrencies}. Further, the advantages offered by the blockchain technology have heralded a growth in its possible applications in many fields \cite{underwood2016blockchain,bodkhe2020blockchain,catalini2020some,holbl2018systematic}. Basically blockchain is an ever-growing list of decentralized records that is distributed between mutually cooperating parties who do not trust each other \cite{kolb2020core}. Further, the data is stored in the forms of blocks which are linked together via the use of cryptographic hash functions. The objective is to have a database containing records of authentic transactions which cannot be tampered even if some of the parties become malicious. Thus, the main idea is to have a tamper-proof database which is not under the control of any central authority with all the competing yet cooperating parties having an underlying stake. The main primitives of the blockchain technology are as follows:

\textbf{A. Digital signature:} In a blockchain, records are considered as a virtual asset which is to be secured against a malicious user. Suppose one user wants to perform a transaction (e.g., sending money to other person or casting the vote for a candidate), then digital signatures provide a mechanism to ensure the authenticity of such transactions. For the purpose of digital signatures, every user of the blockchain will generate a pair of public and private key which are linked together via a mathematical function by the use of cryptographic protocols, such as RSA-2048, DSA or EC-DSA. The public key will be available to everyone while the user will sign transactions by using the private key.  When the transaction is sent to the pool of unconfirmed transactions then the public key will be used to verify the authenticity of the transaction else it will be discarded. Hence, the digital signatures serve the purpose of authentication of the genuine transactions.

\textbf{B. Hashing:} Hashing is basically a one way trapdoor function in which a cryptographic algorithm is used to generate a hash of the input data. The interesting part of hashing is that the hash output is a random string of fixed length irrespective of the amount of input data. Even if there is a change in one bit in the input data, the corresponding hash will be considerably different and cannot be predicted beforehand. Further, it is not invertible, i.e., if one knows the hash then it is hard to get the corresponding input data. Hashing is used to make the records in blockchain tamper-proof. The hash of the data present in the previous block is stored in the current block. Therefore, if a malicious user wants to tamper any data in the $n^{th}$ block then the corresponding hash value will change drastically, and the malicious user has to make appropriate changes in the successive blocks from $n+1^{th}$ block onward. Thus,  a malicious user has an uphill battle to get all these blocks accepted by the miners, which is practically not possible  unless  he/she controls more than 50 percent of the computational resources. Hence, hashing provides a way to link the successive blocks in the blockchain to each other and thus making it tamper-proof.

\textbf{C. Miners:} In the decentralized system, miners are basically users who are responsible for verifying the transactions from the pool of unconfirmed transactions, combining them in the blocks and finally adding the block to the blockchain. Anyone can become a miner, but for getting the rights for adding a block, the miner has to fulfil certain conditions as per the consensus protocol. For instance, in Bitcoin, a miner  has such privilege if he solves a computationally hard problem of finding the hash value of the new block which is less than the specified threshold value of hash. Further, the contents of the new block can be verified by any miner, and the block will not be further extended if it fails in the validation check; and  such a block will be called as an orphan block. Moreover, all the miners also keep a local copy of blockchain with themselves.

\textbf{D. Consensus protocols:} They  provide a mechanism via which miners decide to agree on the addition of blocks in the blockchain. Popular among the consensus protocols are proof of work (PoW), proof of stake (PoS), delegated proof of stake (DPoS).  In PoW, a user is granted access as a miner to add the blocks only after solving a computationally hard problem. Thus, the miners have to utilize his computational resources in order to become eligible to add a block. This leads to a system in which there is some considerable gap between the successive  addition of the blocks. For example, in Bitcoin, it takes an average of 10 minutes to add a block. This in a way  prevents any particular miner or a group of miners to indiscriminately add new blocks and hence monopolize the blockchain.  In PoS, the miners get the rights to add the block in proportion to their stakes in the blockchain. While in DPoS, all the miners from the current set of available miners take turns in addition of the block. If a miner is not available during his  turn then the  mining right goes to the next available miner. 

Further, the blockchains can be categorized into the public and permissioned blockchains. Public blockchain is suitable where everything is transparent as anyone can become a miner or can check the validity of transactions. However, public blockchains are very slow with regards to the number of transactions, e.g., Bitcoin can handle 3-7 transactions per second which is much less in comparison to  thousands of transactions per second handled by Visa \cite{hazari2019parallel}. In contrast, permissioned blockchains are fast but only selective users { possess} the rights to add the blocks while the rest { have} read only permissions. All the above features of blockchain make it a perfect candidate to implement electronic voting systems with the minimal need of a trusted central authority. Wang et al. \cite{wang2018large} in 2018 proposed a scheme for large scale elections based on blockchain with the blockchain being used just as the storage medium for the voting data. Subsequently, a survey of research work on the implementation of the voting protocols using the blockchain technology was reported \cite{abuidris2019survey}. A comprehensive review of the voting schemes based on blockchain technology with regards to voting requirements both large and small scale at the national elections can be found in  \cite{huang2021application}. In fact, recently there has been a flurry of papers for application of blockchain in electronic voting systems \cite{rathee2021design,abuidris2021secure,jafar2021blockchain,ahn2022implementation,ajao2022application}.

Along with the revolutions provided by blockchain technology in the past decade, the world is at the cusp of another revolution in the field of quantum computation. Even though Shor's algorithm \cite{shor1994} for period finding and Grover's search algorithm \cite{grover1996fast} have been known to everyone since last 30 years but the lack of necessary hardware for a fully functional useful quantum computer precludes their practical utility.  However, things are now getting better day by day with  50 percent probability to  have a fully functional quantum computer by 2031 that can break the RSA-2048 cryptosystem \cite{mosca2018cybersecurity}. Once we have quantum computers in hand then the primitives of classical blockchain technology become vulnerable and hence limit their usefulness in the future. This has been highlighted by Aggrawal et al. \cite{aggarwal2017quantum} in 2016 with regards to potential  attacks on the  crypto currency Bitcoin due to the speed ups provided by any practical quantum computer. Alongside a white paper on the possible ways to make the existing blockchains secure against quantum attacks has been published \cite{gheorghiu2017quantum}. They also proposed newer designs for making blockchains using post quantum cryptographic protocols and compared their usefulness. Further,  Fedorov et al. \cite{fedorov2018quantum} in 2018 commented on the  vulnerabilities  of the current classical blockchain technology using computationally secure cryptography against quantum attacks and suggested ways of incorporating  quantum technologies with blockchain in order to make it quantum safe and offer newer technological innovations. In the same year, the issue posed due to quantum computation on the security and privacy of the blockchain was also addressed \cite{ikeda2018security}. Similarly, Sun et al. \cite{sun2019towards} in 2019 provided a model to remove the limitations of the classical blockchain due to the advantages offered by a quantum computer. They proposed a post-quantum secured protocol for blockchain based on the unconditionally secure digital signature schemes and a new consensus protocol. A brief review of quantum and hybrid quantum/classical blockchain protocols can be found in  \cite{edwards2020review}. It pointed out the issues faced by blockchain, such as scalabilty, efficiency and sustainability. Further, it has been mentioned that quantum computers can pose challenges to the application of classical blockchain in mission critical work. Therefore, it is necessary to integrate the blockchain technology with the quantum computational issues. Some quantum analogue of bitcoins using the power of no-cloning has been introduced but is not yet popular. In the same year, another review of blockchain cryptography resistant to quantum cryptography was published, where some ways to secure blockchain against quantum attacks were proposed \cite{fernandez2020towards}. 

We can see that most of the currently suggested solutions to safeguard the blockchain against quantum attacks is via the use of post-quantum cryptographic protocols. Such protocols are computationally secure for the time being but one does not know of the future. It may be possible that some new quantum algorithms may come up which may make even the post-quantum cryptographic schemes vulnerable. Thus, a natural step is to integrate the unconditionally secure quantum computation based technologies with the blockchain technology in order to make the records eternally safe. A step in this direction was taken by  Kiktenko et al. \cite{kiktenko2018quantum} in 2018 by proposing  a model for quantum blockchain by using the quantum key distribution (QKD) protocols, such as BB84 protocol for communication between the nodes. Further, quantum Byzantine agreement protocol \cite{lamport2019byzantine} via the use of interactive consistency vector is used to generate consensus for addition of new blocks in the system. The protocol is secure if there in no more than one third of the cheating nodes. Also, they have experimentally tested the protocol by means of a three-party urban fibre network QKD. This protocol provided the spark and Sun et al. \cite{sun2019simple} in 2019 proposed a simple voting protocol based on the quantum blockchain. However, this protocol has certain limitations as it cannot be scaled up beyond a certain small number of users. Also, it is very easy for a particular voter to use the denial of service attack to other voters.  With regards to the proposal by  Kiktenko et al. \cite{kiktenko2018quantum}, even though quantum blockchain  seems interesting but a lot has to be improved. First and foremost is to maintain the trade-off between the contrasting assumptions of QKD and blockchain technology as QKD is based on the assumption that nodes are authenticated, but blockchain does not require the nodes to authenticate each other. Also, the quantum Byzantine agreement protocol becomes exponentially very data intensive under the presence of a large number of cheating nodes. Other than that, a conceptual design of fully quantum blockchain was proposed via use of GHZ states state of photons which are entangled in time with each other \cite{rajan2019quantum}. Further, Gao et al. \cite{gao2020novel} extended this idea of entanglement in time to propose a novel blockchain scheme and introduced quantum version of crypto-currency; safety of which is guaranteed by the no cloning theorem. In the same year, Banerjee et al. \cite{banerjee2020quantum} proposed to build a quantum blockchain using the multi-party entanglement of quantum weighted hypergraph states and provided its quantum circuit and proof of principle implementation on IBM’s five-qubit quantum computer. Even though the above works provide only toy models for implementation of secure quantum blockchains but a study in this direction is necessary to integrate the two technologies. Further, it would be nice to develop these technologies with an application bent of mind, with secure electronic voting being one of them. 

In the next two sections,  we will propose a protocol for remote electronic voting scheme by integrating the blockchain technology with the quantum enabled technologies, such a QKD schemes and quantum random number generator (QRNG). The idea is to combine their respective advantages and propose solutions which can be implemented using the currently available resources.

\section{Quantum Resources}

Before going on to the details of the voting system, let us just briefly describe two of the most important quantum  resources required for the implementation of the scheme.

\subsection{Quantum Random Number Generator}

Random numbers are one of the most important requirements for many of the practical tasks, such as lotteries, Monte Carlo simulations, statistical procedures. Currently, there are many classical algorithms that efficiently generate random numbers at will \cite{lehmer1951mathematical,matsumoto1998mersenne}. However, the random numbers generated through the algorithmic processes have a disadvantage that the generated random numbers have a periodicity (maybe be very large) and hence are called as pseudo-random numbers. Further, any adversary who knows the algorithm may predict the future outcomes after getting hold of some of the generated outputs. Hence, they are not suitable for applications which require genuine randomness. In comparison to algorithmic  processes, quantum systems have intrinsic randomness \cite{bera2017randomness} and provide a mechanism to generate true random numbers, e.g., the radioactive decay is a quantum phenomenon where the decay is decided randomly, noise present in the electronic circuits can be used to generate true random numbers. One of most easiest ways to generate true random numbers is via the use of optical devices \cite{ma2016quantum}, e.g., the photon from a single photon source is allowed to fall on a beamspliter and detectors are placed on both the arms. Due to the quantum behaviour of the photons only one of the detectors will click, and it will be decided randomly. Hence, such a simple scheme can be used to generate a true random number sequence. Currently, there are various ways via which the true random numbers can be generated using the optical devices as well as other implementations with many commercially available products too \cite{herrero2017quantum,mannalath2022comprehensive}. The generation rates of the random numbers can vary for the devices but it can go to the order of Gbps. Such true random number generators can be used in applications which require genuine randomness. 

Even though the quantum random number generators produce true random numbers, one has to trust the vendor who is manufacturing the device and such devices are known as trusted random number generators.  However, this is not a valid assumption in the realistic scenarios. In practice, the device may be defected or could even be in the control of an adversary, and this can have serious ramifications of the process where such a device is used. Therefore, one has to verify that the output generated from the devices are indeed truly random in nature. For this purpose, we can have self testing quantum random number generators, in which the user can certify the quantum random process by the inspection of input-output statistics \cite{colbeck2009quantum}. Basically, the self testing quantum random number generators use the violation of Bell's inequality (or any of its variants) to certify the randomness of the devices \cite{Pironio_2013,miller2015universal,Gallego_2013}. Such self testing QRNGs have a very  high   credibility but have very low generation rates in comparison to the trusted device QRNGs. Interestingly, we can have semi self testing QRNGs which have good credibility as well as better generation rates \cite{Cao_2016,Marangon_2017}. Currently, there are many off the shelf commercial trusted QRNGs \cite{jacak2021quantum,mannalath2022comprehensive}, such as ID Quantique, Toshiba, PicoQuant, qStream with excellent generation speeds, while the self testing and semi self testing devices are still  at the experimental stages.

\subsection{Quantum Key Distribution}

We all know that cryptography is the art of sending secret messages, and the secret is secure as long as the key for encoding the message is secure. In view of this, the problem of secure transmission of message can be considered as the problem for secure distribution of the secret keys. Public key cryptography \cite{hellman2002overview} provided a revolution in this field by solving this problem, where the message is encoded by the public key of the receiver and message is decoded by the help of the private key of the receiver. Even though the public key of the receiver is available with everyone but getting the private key is computationally impossible for any adversary \cite{rivest1978method}. Thus, the public key cryptography is computationally secure. However, Shor \cite{shor1994} in 1994 showed that  the RSA algorithm can be broken if one has a quantum computer in hand. This led to a renewed interest towards the development of unconditionally secure key distribution. In comparison to the computational security provided by the classical cryptography, the keys generated by the quantum cryptography are unconditionally secure. This field of unconditionally secure quantum cryptography commenced with the first QKD protocol proposed by Bennett and Brassard in 1984, now known as BB84 protocol \cite{bennett1984quantum}. Later on several new quantum protocols were introduced and experimentally realised \cite{gisin2002quantum} which can be classified in the schemes, such as entanglement based protocols, orthogonal state based protocols, semi-quantum protocols \cite{shenoy2017quantum}. In particular, the entanglement based QKD introduced by Ekert \cite{ekert1991} is of the special interest as it later formed the basic idea of device independent secure key distribution \cite{vazirani2019fully}. From the commercial aspect, QPN security gateway (QPN-8505) proposed by MagiQ Technologies was the first commercial QKD system \cite{korchenko2010modern}. Even though the  key rates and the communication distance were not great but it provided a new spark. The current technological revolution has now led to newer companies, such as ID Quantique, Toshiba, QinetiQ offering secure QKD solutions. Every year, the maximal distance and key rates are growing \cite{xu2020secure}. QKD will surely become a norm for practical high security transmissions  sooner-or-later.

\section{Voting scheme based on quantum assisted blockchain}

 The permissioned quantum blockchain forms the backbone of the proposed voting scheme. In this scheme, every user has the permission to read the contents of the blockchain, but only some selected persons are allowed to become miners for addition of a block. The proposed voting scheme consists of the following major stakeholders who are responsible for the successful conduct of elections. \\

\begin{figure}[h]
\includegraphics[width=\textwidth]{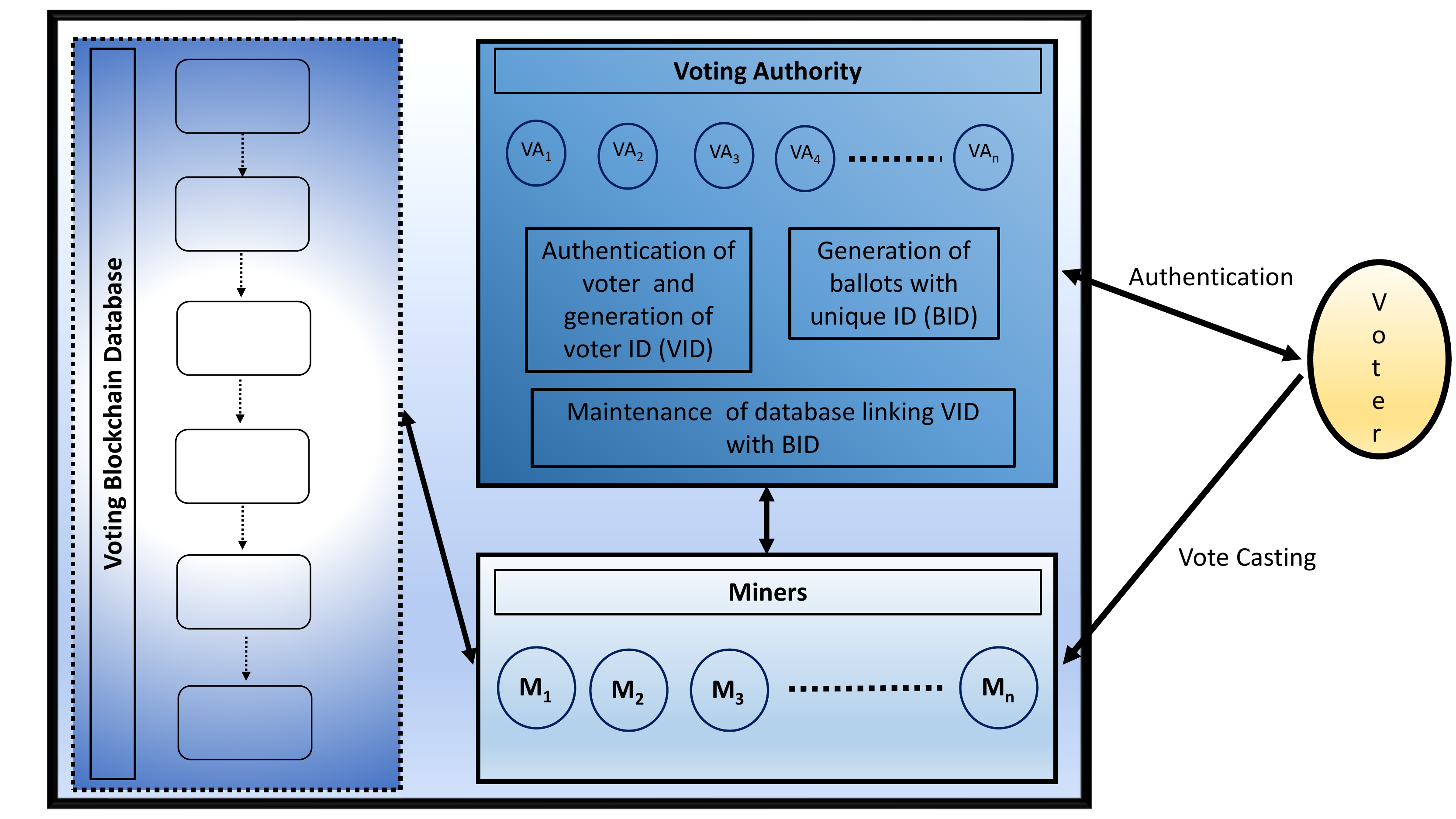}
\caption{ Schematic architecture of the electronic voting scheme using quantum assisted blockchain. } 
\label{fig1}
\end{figure}

\textbf{Voting Authority} 

Voting authority ($VA$) has the overall responsibility for the conduct of election, and it will consist of multiple number of personnel with the power being shared between them. The sharing of power among the personnel removes the possibility of manipulating the outcome of the voting by a few of the dishonest members. The main role of the voting authority is to take care of the registration of voters by checking their credentials and to form  a pool of a set of voters ($\mathcal{V}$), who are eligible to cast their votes in the elections. A random unique Id is provided to each voter from the voter pool ($\mathcal{V}$), and  $VA$ records this information for the purpose of authentication   to commence the voting process. Further this is useful for other processes, such as tallying and audit of the elections. $VA$ have the full responsibility to delegate the powers to the people who will work as the miners for the addition of votes into the blockchain. Also, $VA$ will interact with the miners for the purpose of authentication of the votes. Finally, the result of the elections will be declared by $VA$ after completing the audit of the votes.

\textbf{Voters} 

Voters are central to any scheme of voting. In order to become a part of the eligible pool of voters ($\mathcal{V}$), every voter has to get registered with $VA$ before the casting of the votes. Voters will cast their votes as per the rules set by $VA$ by sending their votes to the miners. Further, the voters can verify their votes by looking at the set of records present in the voting blockchain.

\textbf{Blockchain and miners} 

Miners ($\mathcal{M}$) have the overall responsibility for the upkeep of the voting blockchain database. They will receive the votes from the voters and will   authenticate the votes   with the help of the voting authority. After verification of the vote it is finally added to the voting blockchain by making use of DPoS protocol. In this protocol, $'n'$ miners will be selected by $VA$ by taking care of the respective representations  from the candidate pool, voting pool, civil society and independent observers.  All the miners will take turns while verifying the votes from the set of unconfirmed pool of votes and after checking their validity will add the block to the voting blockchain. If at any stage, the miner who is assigned the task is not available, then that miner will loose the turn and mining power will be transferred to the next available miner. Further, a block will be considered as a valid block only if the majority of the miners validate its contents. In this way, there will be a transparent mechanism by which voting data is recorded in the voting blockchain. \\

The voting process of the proposed protocol involves the following stages:

\subsection{Pre-Voting }

The steps involved in this process are as follows:

\begin{description}

\item[Step:1] Voters sends their credentials to the voting authority via a classical authenticated channel. We can use the quantum channel in this step if the number of voters is small. Since the cost of using quantum channel is very high, it is always advisable to use a classical authenticated channel for large elections (e.g., a channel authenticated by  Wegnam-Carter authentication scheme \cite{wegman1981new,carter1979universal}).

\item[Step:2] Voting authority ($VA$) verifies the credentials of the voters as per his/her own records or via links to any trusted third party. After verifying the credentials of the voter, voting authority uses a quantum random number generator to generate a sufficiently long (256 bit) unique Id (VID) for the voters. The purpose of generating a random unique Id for every voter is to implement the anonymity of the vote.  Further, VID and its hash value will help in the auditing of the voting process without revealing the identity of the voter.

\item[Step:3]  VID along with its hash value is stored at the database of the voting authority in an encrypted way. For the security of the database, encryption and decryption keys are split  between multiple members working in the voting authority via the well-known quantum secret sharing protocols \cite{hillery1999quantum,xiao2004efficient,tittel2001experimental}. 

\item[Step:4] The random unique Id (VID) along with the signature of the voting authority will be sent to the voter in an encrypted manner. For the large elections or voting of the lesser importance, one can use public key cryptography. However, for a small number of voters or crucial elections, only BB84 or a practical variant of it will be used to transfer VID from $VA$ to the voters.

\end{description}

\subsection{Voting Process}

The steps involved in this process are as follows:

\begin{description}

\item[Step:1] Voting authority prepares the ballot by generating a unique long (256 bit) ballot Id (BID) using a quantum random number generator for every voter. Thus, for every VID, there will be a corresponding BID. 

\item[Step:2] Voting Authority stores the pair of VID and BID for every voter in the voting database in an encrypted manner using an available quantum secret sharing protocol. Further, the hash value of the pair of VID and BID is publicly announced.

\item[Step:3] Each Ballot is sent to the corresponding voters in the encrypted manner with the mechanism similar to that  during the transmission of the VIDs to voters.

\item[Step:4] Voters will be sending their votes to any of the authorized miners to form a list of unconfirmed votes. The structure of the vote includes: VID, BID, voting choice, time stamping, voter's signature.

\item[Step:5] The miners will take turns in adding block to the blockchain by following DPoS protocol. 

\item[Step:6] The miner will check for the authentication of the voters by checking for  voter Id (VID) and  ballot Id (BID) combination  under the supervision of the voting authority. If the vote is genuine, then the hash value of the VID and BID pair will match with that announced by the voting authority. After the successful authentication, the vote will be added to the block. Further, the block will be added to the voting blockchain. 

\item[Step:7] For the purpose of security, the miners will interact  among themselves and with the voting authority  using QKD secure keys produced in BB84 protocol or one of its variants. A block will be considered as a valid block only if the majority of miners validate its contents.  \\ 

Since the voting is taking place via remote login protocol, there may be situations in which the voter may vote under undue pressure or influence form an external agent who wants to rig the elections. An appropriate mechanism for security against such situations is to grant a procedure by which such voter may cast his vote afresh. The steps involved in the recasting of the votes are as follows:   

\item[Step:A] Voter will again request the voting authority to provide a new ballot.

\item[Step:B] Voting authority will allot a new ballot ID and discard the  previous ballot Id associated with the voter. Further, the new ballot Id corresponding to the voter's unique ID will be stored in the voting database. This new ballot Id will be sent to the voter in a secure manner as described earlier. 

\item[Step:C] Voter will vote with the new ballot Id and submit the vote to the miner. The miner will add the vote to the blockchain after verifying the validity of vote. \\

\item[Step:8] The voting process terminates at a pre-scheduled time. Thus, only votes received till the cut-off time will be considered to be included into the blockchain.

\end{description}

\subsection{Tallying Process}

After the conclusion of the voting process, the voting blockchain data will be available for public viewing. Votes can be tallied by any miner or any voter as the blockchain contains the record of all the valid votes.  All the valid votes (having correct pair of VID and BID) can be verified by anyone. For the case of multiple votes with the same pair of VID and BID, only the vote that is cast earliest will be counted. The voting authority announces the result, and the results can be verified by all the stakeholders.  

In the following, we will look at the security of the proposed voting scheme.

\section{Security}
Any proposed protocol of the voting cannot be considered as a valid scheme until and unless it conforms to the requirements of voting. In this section, we will show that our proposed protocol is a valid scheme for voting as it satisfies all the requirements of the security. This is described as follows:

 \subsection{Eligibility}
 
 The first and foremost requirement in any voting scheme is to make sure that only the eligible voters are allowed to take part in the voting process. For this purpose, every voter will have to register with the voting authority before the deadline. The voting authority will follow the due process to verify the credentials of the applicant voters. Voting authority will then form a pool of the eligible voters who can take part in the voting. Only the voters who are eligible will be provided voting ID (VID). Therefore, only the voters with valid  VID can further take part in the voting process.  To commence the voting, all the voters with a valid VID will be provided a randomly generated 256 bit BID. The miners will add only those votes in the blockchain which will have the valid pair of VID and BID. Since the VID and BID are generated by a QRNG so it will be very difficult for an adversary to guess the correct pair of VID and BID. In this way, only eligible voters with the valid pair of VID and BID can vote, otherwise their votes will be discarded by the miner. Thus, the eligibility condition is fully satisfied in the proposed scheme.
 
\subsection{ Anonymity}

The anonymity of the vote is very important in voting as voters may have to face some repercussions if the voting choice is revealed.  In order to maintain the anonymity of the vote, all the eligible voters are provided a random VID generated by a QRNG.  The voters are again allotted a randomly generated BID on the voting day. The voters then cast their votes by sending the votes to miners which included VID, BID, voting choice, time stamping, and voter's signature. It is to be mentioned here that the signature used by the voter at the time of casting the vote is linked to VID only rather than the real identity of the voter. Thus, the contents of the blockchain does not, in any way, reveal the identity of the voters as it only contains the valid pair of VID and BID along with the  voting choice.   Also, the anonymity of the vote is maintained. It is worth mentioning here that voting authority maintains a record of valid VID and BID. Thus, he/she can, in principle, know how the voter has voted. However, the contents of database containing the record of eligible voters along with the valid VID and BID is stored in an encrypted manner with the use of quantum secret sharing schemes.  Since the key is shared between multiple users containing people from various fields, such as voters, candidates, civil society, and others, so only if they all agree, no one will be able to reveal the anonymity of the vote. Hence, the anonymity condition is satisfied.

\subsection{ Non-reusability}

In an election, mechanism should be there so that a voter cannot vote more than once. In this scheme, all the voters are provided a valid pair of VID and BID which can be used only once while casting the vote. Whenever the voter sends the vote to the miner, then the miner checks for the validity of the vote, and the vote is added in the blockchain only after verification. It may happen that the voter can send vote to different miners in the hope of adding multiple votes. However,  note that for a valid pair of VID and BID, only the vote which has been cast first will be counted in the tallying phase and the rest will be discarded. Therefore, if at all the blockchain contains multiple entries of votes with valid pair of VID and BID, then only the vote that has been cast earliest will be counted. In this way, only a single vote for the eligible voter will be counted.    

\subsection{Binding}

An important property of a good voting system is that no one including the voter can alter the vote once it has been recorded. The voters use the VID and BID issued to them for casting the votes. The validity of the votes can be identified by matching with the hash values of all the eligible pairs of VID and BID. The vote is then added to the voting blockchain by the miners. As per the property of the blockchain, the contents cannot be altered by anyone as the blocks are connected to each other via cryptographic hash functions. Thus, the votes, after it has been recorded by the blockchain, will remain intact and hence the binding property is satisfied.     

\subsection{Verifiability}

The voting blockchain data is  publicly accessible to everyone after all the eligible votes have been recorded. Thus, any voter who wants to verify can look into the voting blockchain and verify whether his/her vote has been properly recorded or not. Further, the results announced after the tallying phase can be verified by any voter as the voting blockchain is made available to anyone. This ensures verifiability condition of the proposed scheme.  

 \subsection{Auditable}
 
 The main advantage of the proposed scheme is that it is very easy to undertake the audit of every process. Since the voting blockchain data is available publicly, so it is very easy for audit by any independent authority without revealing the identity of the voters. Further, the voting authority and the miners contain multiple persons who represent various stakeholders of the elections which makes the voting process very transparent. Therefore, the scheme is transparent and auditable.

\section{Discussions and Conclusion}

All democratic institutions use voting as one of the principle tools to arrive at a decision. But as mentioned before, many times it is important to preserve the anonymity of the vote so that voters can vote as per their freewill. Mankind has been working tirelessly to develop systems that conform to the requirements of a good anonymous voting system.  In fact, voting forms a sub-field of cryptography known as secure multi-party computations in which multiple parties try to compute a joint function over inputs without revealing the individual inputs \cite{canetti1996adaptively}. Since the advent of quantum cryptographic protocols \cite{gisin2002quantum,xu2020secure} to generate unconditionally secure  keys, researchers are continuously working on the related problems of quantum secure multi-party computations \cite{colbeck2009quantum}, such as quantum voting \cite{vaccaro2007quantum,hillery2006towards,thapliyal2017qv}, quantum auction \cite{naseri2009secure,sharma2017quantum}, private comparison \cite{yang2009secure,thapliyal2018orthogonal}, quantum lottery \cite{mishra2022lottery}, in order to develop realistic unconditionally secure protocols. The inherent properties of quantum system provide a perfect platform to propose solutions to the mentioned challenges. For the case of voting, quantum solutions were first proposed by Vaccaro et al. \cite{vaccaro2007quantum} and Hillery et al. \cite{hillery2006towards} in 2006. Since then, the field of quantum voting is flooded with a flurry of papers \cite{thapliyal2017qv,wang2016qv,jiang2012qv,xue2017qv,mishra2022quantum}. However, fully quantum solutions of voting have their own set of challenges.  With the current level of advancements, except the solutions in the field of quantum cryptography and quantum random number generator, the quantum solutions for other areas are still  at the nascent stage to be used for practical and commercial applications. The reason being that the quantum states are very fragile, and they decohere under the influence of environment. Further, the quantum states to be used in the solutions for voting are often too complex to be realised and maintained under the realistic situations. On the other hand, such quantum solutions fail to provide a scale up as the number of voters starts to increase. Hence, the proposed quantum voting protocols are at this moment are technologically relevant only for the case of limited number of voters and that too under the presence of highly sophisticated devices used for the implementations. 
 
As mentioned in the previous sections, blockchain technology has now emerged as the technology  with the potential to revolutionize many fields of life. Electronic voting is one such application for which blockchain seems to be a perfect candidate. Specifically, for the commercial development of the solutions of problems of interest with a host of protocols and implementations coming in the last few years \cite{wang2018large,abuidris2019survey,huang2021application}. However, the proposals to implement electronic and internet voting based on blockchain have also faced some criticisms \cite{park2021}. The main concerns were that the current blockchain based systems are still far-off from being perfect to detect errors, malfunctioning and malware. Moreover, use of blockchain technology can open up new windows for attack. Further, the cost of altering the results by an adversary is substantially lesser than that for the case of a traditional paper ballot voting system. This can lead to a huge security risk on elections of high importance if it is to be implemented using the current blockchain technology.  Therefore, we can see that although the current blockchain based electronic voting implementations can be fairly deployed for elections where the stakes are low, an elaborate security audit is necessary before being deployed for the high-stakes elections. 

In this work, we have carefully reviewed the characteristic features of the current blockchain technology with an emphasis on the applications in the field of secure electronic voting systems. The idea is to highlight the  advantages and limitations of blockchain more specifically with respect to the upcoming challenges posed by noisy intermediate scale quantum (NISQ) computer. Though NISQ computer is still to be realized, but one has to develop the electronic voting systems which are full-proof against the possible attacks. Further, it is very important to address the issues raised against blockchain based voting systems. Thus, our work is a step in this direction to enhance the security aspects by carefully utilizing and combining the advantages offered by quantum technology and the blockchain. Here, we have proposed an electronic voting scheme based on quantum assisted blockchain. The permissioned blockchain forms the the backbone of the proposed scheme which acts as the storage medium for the votes, while the addition of quantum technologies, such as quantum random number generators, QKD and QSS protocols, makes the voting systems secure against any adversary. Although, we have retained the concept of the central authority for a smooth conduct of the elections, the powers have been distributed among different set of persons in such a way that it is almost next to impossible for a small group of people to adversely affect the voting outcome. Further, we have shown that the proposed scheme meets all the requirements of a good voting scheme.  It is also worth mentioning here that we have addressed the issues at a higher level only without going too much into the details of challenges to be faced at the implementation stage. The proposed anonymous electronic voting scheme via the use quantum assisted blockchain can be used for the implementation of the large scale elections using the currently available technology. We hope that the present work will provide crucial insights towards the ultimate task of developing an electronic voting system that can be used for elections of high stakes too.

\begin{acks}
SM, APar and APat  acknowledge the support received through the  project “Partnership 2020: Leveraging US-India Cooperation in Higher Education to Harness Economic Opportunities and Innovation” which is enabling a collaboration between University of Nebraska at Omaha, and JIIT, Noida. KT acknowledges support from ERDF/ESF project ‘Nanotechnologies for Future’ (CZ.02.1.01/0.0/0.0/16\_019/0000754).
\end{acks}

\bibliographystyle{ACM-Reference-Format}
\bibliography{VQB}

\end{document}